\documentclass[aps,showpacs,nofootinbib,amsmath,twocolumn, tightenlines]{revtex4}
\usepackage{graphicx}

\usepackage{color}
\begin{document}

\title{\boldmath
 Spin correlations in the reaction  ${ {\pi}^\pm \vec D \to\vec\Sigma^\pm\Theta^+}$
  and the parity of $\Theta^+$}

 \author{A.\,I.~Titov$^{a,b}$ and B. K\"ampfer$^{a}$}

 \affiliation{
 $^a$Forschungzentrum Rossendorf, 01314 Dresden, Germany\\
 $^b$Bogoliubov Laboratory of Theoretical Physics, JINR,
  Dubna 141980, Russia\\
  }

%%%%%%%%%%%%%%%%%%%% Abstract %%%%%%%%%%%%%%%%%%%%%
\begin{abstract}
We analyze  two types of spin observables in the reaction $\pi
\vec D\to\vec\Sigma\Theta^+$ near the threshold. One concerns the
spin-transfer coefficients $K_x^x$ and $K_z^z$. The second one is
the deuteron spin anisotropy.  These observables  are sensitive to
the $\Theta^+$ parity and can be used as a tool for the $\Theta^+$
parity determination.
\end{abstract}


\maketitle

\section{Introduction}

 The first evidence for the  pentaquark particle  $\Theta^+$ discovered by the
 LEPS collaboration at SPring-8~\cite{Nakano03} was subsequently
 confirmed in other experiments~\cite{OtherPenta}. However
 some other experiments fail to find the $\Theta^+$ signal
 (for a review see~\cite{Hicks}).
 None of the experiments with positive signal of $\Theta^+$
 determined the spin and the parity of
 the $\Theta^+$. Since most of evidences
 of the pentaquark came from photoproduction,
 then naturally there were several proposals
 for determination of the parity of $\Theta^+$, $\pi_\Theta$,  from photoproduction
 processes using single and double polarization
 observables. However, these suggestions
 are based on model-dependent calculations.
 Only triple spin observables may be considered as
 reliable candidates for a determination of $\pi_\Theta$, but the practical
 implementation of this method is rather hard (for a review see~\cite{TEHN04}).
 Therefore, one should look for other
 reactions and observables for  elucidating $\pi_\Theta$,
 since the determination of $\pi_\Theta$ is challenging to
 pin down the nature of $\Theta^+$

 In Ref.~\cite{THH}, Thomas, Hicks and Hosaka have proposed an
 attractive method to unambiguously determine the $\Theta^+$ parity
 in the reaction $\vec p\vec p\to \Sigma^+\Theta^+$
 close to the production threshold. They have shown that the
 initial proton-proton spin state with spin $S=0\,(1)$ is
 completely defined by the
 the $\Theta^+$ parity ($\pi_\Theta=+(-)$). The method is solely based on the
 Pauli exclusion principle and the total spin and parity conservation as
 well, and therefore is model independent, indeed. This idea
 with utilizing other double spin observables (spin-spin correlations)
 in $NN\to Y\Theta^+$ reactions
 has been further developed by Hanhart et al.~\cite{Hanhart1,Hanhart2},
 Rekalo and Tomasi-Gustafsson~\cite{Rekalo1,Rekalo2},
 Uzikov~\cite{Uzikov1,Uzikov2}, and Nam, Hosaka and Kim~\cite{Nam}.
 In particular, the focus was put on an analysis of the
 spin-transfer coefficients which are sensitive to the production
 mechanism and $\Theta^+$ parity.
 It seems to be interesting and
 important to elaborate further and alternative methods for
 an unambiguous determination of $\pi_\Theta$
 which serve an independent check of internal $\Theta^+$ properties.

 In our Communication we consider
 the reaction $\pi^\pm \vec D\to \vec\Sigma^{\pm}\Theta^+$ near the
 threshold.
 We analyze (i) the spin-transfer coefficients $K_i^i\,
 (i=x,y,z)$,
 where the spin is transferred  from the polarized deuteron to
 the outgoing $\Sigma$,
 and (ii) the deuteron spin anisotropy ${\cal A}$, which defines the production
 cross section as a function of the angle between the deuteron spin
 and direction of the pion beam. The latter observable has
 an obvious advantage because it bases on single polarization (polarized deuteron
 target) measurements.

 For the sake of clarity, in the following we limit our
 discussion for determining the $\Theta^+$ parity to an isoscalar
 spin-1/2 $\Theta^+$. In fact, most theories predict  $J^P$ of
 $\Theta^+$ to be $1/2^+$ or $1/2^-$. The generalization for
 higher spin of $\Theta^+$ is straightforward.
 Our consideration
 resembles the previous works~\cite{Hanhart1,Uzikov1} on
 $NN$ reactions. The main difference is
 that in our case the total isospin and the spin of the
 $np$ system (deuteron) in the initial state are fixed.
 The deuteron polarization with respect to the beam direction
 provides additional observables compared to the $NN$ reaction.

 \section{Spin-transfer coefficients}

Spin-transfer coefficients for the reaction $\pi  \vec D\to\vec
\Sigma \Theta^+$\footnote{In what follows, the charge indicating
superscript $(\pm)$ will be suppressed for $\pi^\pm$ and
$\Sigma^\pm$.} are related to the production amplitude $T$
as~\cite{Bilenky}
 \begin{eqnarray}
 K_i^f=\frac{{\rm Tr}[T\,S_i(D)T^\dagger\sigma_f(\Sigma)]}
       {{\rm Tr}[T\,T^\dagger]}~,
 \label{E1}
 \end{eqnarray}
 where $\sigma_f$ and $S_i$ are the Pauli spin-$\frac12$
 and spin-1 spin matrices, respectively. The latter ones are defined as
{\small
\begin{eqnarray}
 &&S_x=\frac{1}{\sqrt{2}}
 \left( \begin{array}{ccc}
 0 & 1 & 0 \\
 1 & 0 & 1 \\
 0 & 1 & 0 \end{array} \right),\,\,\,
 S_y=\frac{1}{\sqrt{2}}
 \left( \begin{array}{ccc}
 0 & -i &  0 \\
 i &  0 & -i \\
 0 &  i &  0 \end{array} \right),\nonumber\\
 &&\qquad\qquad\qquad S_z=
 \left( \begin{array}{ccc}
 1 & 0 &  0 \\
 0 & 0 &  0 \\
 0 & 0 & -1 \end{array} \right)~.
 \label{E2}
\end{eqnarray}
}
 In the near-threshold region we assume that the final states with
 the orbital momentum $L_f>0$ are suppressed and, therefore, the production
 amplitude has the following form
 \begin{eqnarray}
 T_{m,m';M_D}&=&\sum_{JM,LM_L}
 \langle\frac12m\frac12m'|JM\rangle\,
 \langle 1M_DLM_L|JM\rangle\nonumber\\
 &&\qquad\qquad \times Y_{LM_L}(\hat{\bf k})a^J_L~,
 \label{E3}
 \end{eqnarray}
 where $\hat{\bf k}$ is direction of the pion beam momentum
 with respect to the quantization axis, $m$ and $m'$ are the spin projections
 of $\Sigma$ and $\Theta^+$, respectively, $L$ is the orbital
 momentum in the initial state, $J$ stands for the total angular momentum,
 and $a^J_L$ denotes the partial amplitude.

 For \underline{positive $\pi_\Theta$} the orbital momentum in the initial
 state must be $L=1$ which results in $J=0,1$. The spin-transfer
 coefficients for the case when the quantization axis is taken along
 the beam direction read
 \begin{subequations}
\label{E4}
 \begin{eqnarray}
 K^z_z&=&\frac{3r^2}{1 + 3r^2}~,\label{E4-1}\\
 K^x_x&=&K_y^y=\frac{\sqrt{6}{\alpha r}}{1+3r^2}\label{E4-2}
 \end{eqnarray}
 with
 \begin{eqnarray}
 a^0_1&=&|a^0_1|\,{\rm e}^{i\phi^0_1},\,\,
 a^1_1=|a^1_1|\,{\rm e}^{i\phi^1_1},\,\,
 r=|a^1_1/a^0_1|,\nonumber\\
 \alpha&=&\cos\delta^+\equiv\cos(\phi^0_1-\phi^1_1)~.
 \nonumber
 \end{eqnarray}
 \end{subequations}
  Taking the ratio $r$ from Eq.~(\ref{E4-1}),
 \begin{eqnarray}
 r=\sqrt{\frac{K_z^z}{3(1-K^z_z)}}~,
 \label{E5}
 \end{eqnarray}
one can express $K_x^x$ through $K_z^z$ as
\begin{eqnarray}
 K_x^x=\pm \alpha\sqrt{{2{K_z^z}}{(1-K^z_z)}},
 \label{E5-1}
\end{eqnarray}
 and one finds the  constraints for $K_i^i$
 \begin{eqnarray}
 0\leq K_z^z \leq 1 ~,
-|\alpha|\frac{\sqrt{2}}{2} \leq K_x^x \leq
|\alpha|\frac{\sqrt{2}}{2}.
 \label{E6}
 \end{eqnarray}

 For \underline{negative $\pi_\Theta$} the orbital momentum in the initial
 state must be $L=0,2$ and the total angular momentum is  $J=1$. The spin-transfer
 coefficients are expressed through the partial amplitudes
 \begin{subequations}
 \label{E7}
 \begin{eqnarray}
 K^z_z&=&     \frac{2+2\sqrt{2}\beta q  + q^2}{3(1 + q^2)}~,\label{E7-1}\\
 K^x_x&=&K_y^y=\frac{2-\sqrt{2}\beta q - 2q^2  }{3(1 + q^2)}~,\label{E7-2}\\
 \end{eqnarray}
 \end{subequations}
 with
\begin{eqnarray}
 a^1_0&=&|a^1_0|\,{\rm e}^{i\phi^1_0},\,\,
 a^1_2 = |a^1_2|\,{\rm e}^{i\phi^1_2},\,\,
 q=|a^1_2/a^1_0|,\nonumber\\
 \beta&=&\cos\delta^-\equiv\cos(\phi^1_2-\phi^1_0)~.
 \nonumber
 \end{eqnarray}
 Eq.~(\ref{E7-1}) allows to express $q$ through $K_z^z$ and $\beta$
 \begin{eqnarray}
 q_{1,2}=\frac{\sqrt{2}\beta \pm D}{3K_z^z-1},
 \label{E8}
 \end{eqnarray}
 with
 \begin{eqnarray}
 \qquad D=\sqrt{9K_z^z (1-K_z^z) + 2(\beta^2-1)}~,
 \label{E8-1}
 \end{eqnarray}
 giving two solutions for spin-transfer coefficient ${K_x^x}_{1,2}$
 as a function of $K_z^z$ and $\beta$. These solutions are related to each
 other as
 ${K_x^x}_1(K_z^z, \beta)= {K_x^x}_2(K_z^z,-\beta)$
 leading to constraints for $K_i^i$
 \begin{eqnarray}
 -\frac{1}{\sqrt{2}}\frac{3}{\sqrt{8+\beta^2}}&\leq& K_x^x \leq
  \frac{1}{\sqrt{2}}\frac{3}{\sqrt{8+\beta^2}},\nonumber\\
  \frac{3-\sqrt{1+8\beta^2}}{6}&\leq & K_z^z
  \leq\frac{3+\sqrt{1+8\beta^2}}{6}~,
 \label{E9}
 \end{eqnarray}
with $d=\sqrt{9+2\beta^2}$. When $\beta^2\simeq 1$, they  reduce
to
 \begin{eqnarray}
 -\frac{\sqrt{2}}{2}\lesssim K_x^x \lesssim
  \frac{\sqrt{2}}{2},\qquad
  0\lesssim K_z^z\lesssim 1~.
 \label{E10}
 \end{eqnarray}

 Interestingly, the found expressions for the spin-transfer
 coefficients in Eqs.~(\ref{E4}) in terms of the partial amplitudes $a^J_L$
 coincide with the spin-transfer coefficients in the $\vec p p\to\Sigma^+\Theta^+$
 reaction with $T=1$ and negative parity. At the same time, Eqs.~(\ref{E7}) coincide
 with the corresponding predictions for the reaction $\vec N N\to Y\Theta^+$
 with $T=0$ and positive parity~\cite{Uzikov1}, but the
 physical meaning of the corresponding partial amplitudes is quite
 different, because of differences in the initial states and the
 production mechanism.

\begin{figure}[h!]
{%\centering
  \includegraphics[width=.6\columnwidth]{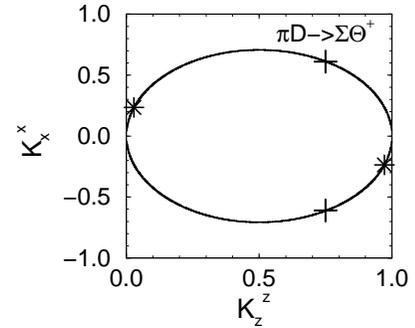}
 \caption{\label{fig:1}{\small%\tcaps%
 The spin-transfer coefficient $K_x^x$
 in the reaction $\pi \vec D \to\vec\Sigma \Theta^+$
 as a function of $K_z^z$ for positive and negative $\Theta^+$ parity
 at $\alpha,\beta=\pm1$.
 The two solutions for $r=q=1$
 for positive and negative $\pi_\Theta$
 are shown by crosses and stars, respectively.}}}
\end{figure}

 Expressions for the spin-transfer coefficient
 are model dependent. They depend on the
 ratios $r$ and $q$ and phases $\alpha$ and $\beta$. However,
 the dependence of one spin-transfer element expressed
 through another one
 is almost model independent and therefore is more attractive.
 As an example, in Fig.~\ref{fig:1} we show
 dependence of $K_x^x$ as a function of $K_z^z$ for
 $\alpha,\beta=\pm1$.

 In this limit the functions $K^x_x(K^z_z)$ do not depend on $\Theta^+$
 parity.
 But in spite of the  similarity in $K^x_x(K^z_z)$  dependence
 one can find a strong difference between spin-transfer coefficients
 as a function of the partial amplitudes. Thus for example,
 if we accept a $democratic$ choice of the partial amplitudes taking
 $r=q=1$, then for the case of positive  parity we find the
 solutions
 $K_z^z\simeq0.75$ and $K_x^x\simeq\pm0.61$. For negative parity
 we have two solutions
 $K_x^x \simeq-0.23$ at $K_z^z\simeq 0.97$
 and $K_x^x \simeq 0.23$ at $K_z^z\simeq 0.03$, for positive
 and negative $\beta$, respectively. These solutions are located
 in the quite different places of the $K_x^x(K_z^z)$ plot,
 as is depicted in Fig.~1.

 We have to note that the correlation $K_x^x(K_z^z)$ depends on the
 phases $\delta^{\pm}$ (or $\alpha,\beta$), and is not
 model-independent. But at the same time, we expect that $|\alpha|$
 and $|\beta|$ are close to one; thus  this dependence is rather weak.

\begin{figure}[h!]
{%\centering
  \includegraphics[width=.7\columnwidth]{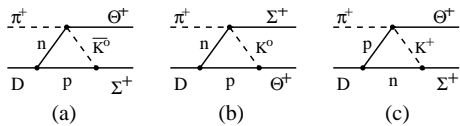}
 \caption{\label{fig:2}{\small%\tcaps%
  Diagrammatic representation of the
 $\pi^+ D \to\Sigma^+ \Theta^+$ reaction. For
  $\pi^- D \to\Sigma^- \Theta^+$ one has to exchange $\pi^+\to
  \pi^-$, $\Sigma^+\to \Sigma^-$ and $n,p\to p,n$.}}}
\end{figure}

\begin{figure}[h!]
{%\centering
  \includegraphics[width=.7\columnwidth]{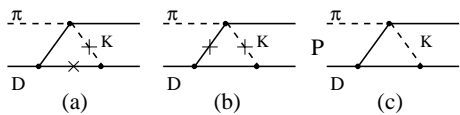}
 \caption{\label{fig:3}{\small%\tcaps%
  Diagrammatic representation of cutting (shown by crosses) the loop
  diagrams (a) and (b);
  (c) corresponds to non-pole background contribution.}}}
\end{figure}

 Indeed, the dominant contribution to the production amplitude
 depicted in Fig.~\ref{fig:2} comes from  the imaginary parts of
 the corresponding diagrams   obtained by cutting  the loops as
 shown schematically in Fig.~\ref{fig:3}a, b and c ($T_p$). The
 non-pole (background) contribution ($T_{bg}$), shown in
 Fig.~\ref{fig:3}d, is much weaker~\cite{gammaD}.
 This means that the phase differences
  $\phi_1^{0} - \phi_1^{1}$
  and  $\phi^{0}_{1} - \phi^{2}_{1}$ are
  close to $\epsilon \pm
 \frac{\pi}{2}$ or $\epsilon +0$ (depending on convention).
 {For positive $\pi_\Theta$, the relative phase
  between the partial amplitudes
  $a^J_{L}$ with the same orbital momentum, $L=1$, and  different
  total angular momenta $J=0,1$ are defined by the spin decomposition
  which does not give an additional imaginary phase between
  $a^0_1$ and $a^1_1$.
  Similarly, for negative $\pi_\Theta$,
  the partial amplitudes $a^1_{L}$ with $L=0,2$
  defined by the angular harmonic decomposition of the
  total amplitude have an orbital phase factor $i^L=\pm 1$
  which does not provide additional imaginary phase
  between $a^1_0$ and $a^1_2$.}
 Therefore, the corresponding relative phases $\delta^{\pm}$ in
Eqs.~(\ref{E4}) and (\ref{E7}) can be estimated as
$|\cot\delta^\pm|\sim |T_p|/|T_{bg}|$ being close to 0 or $\pi$.

\begin{figure}[t!]
{%\centering
  \includegraphics[width=.45\columnwidth]{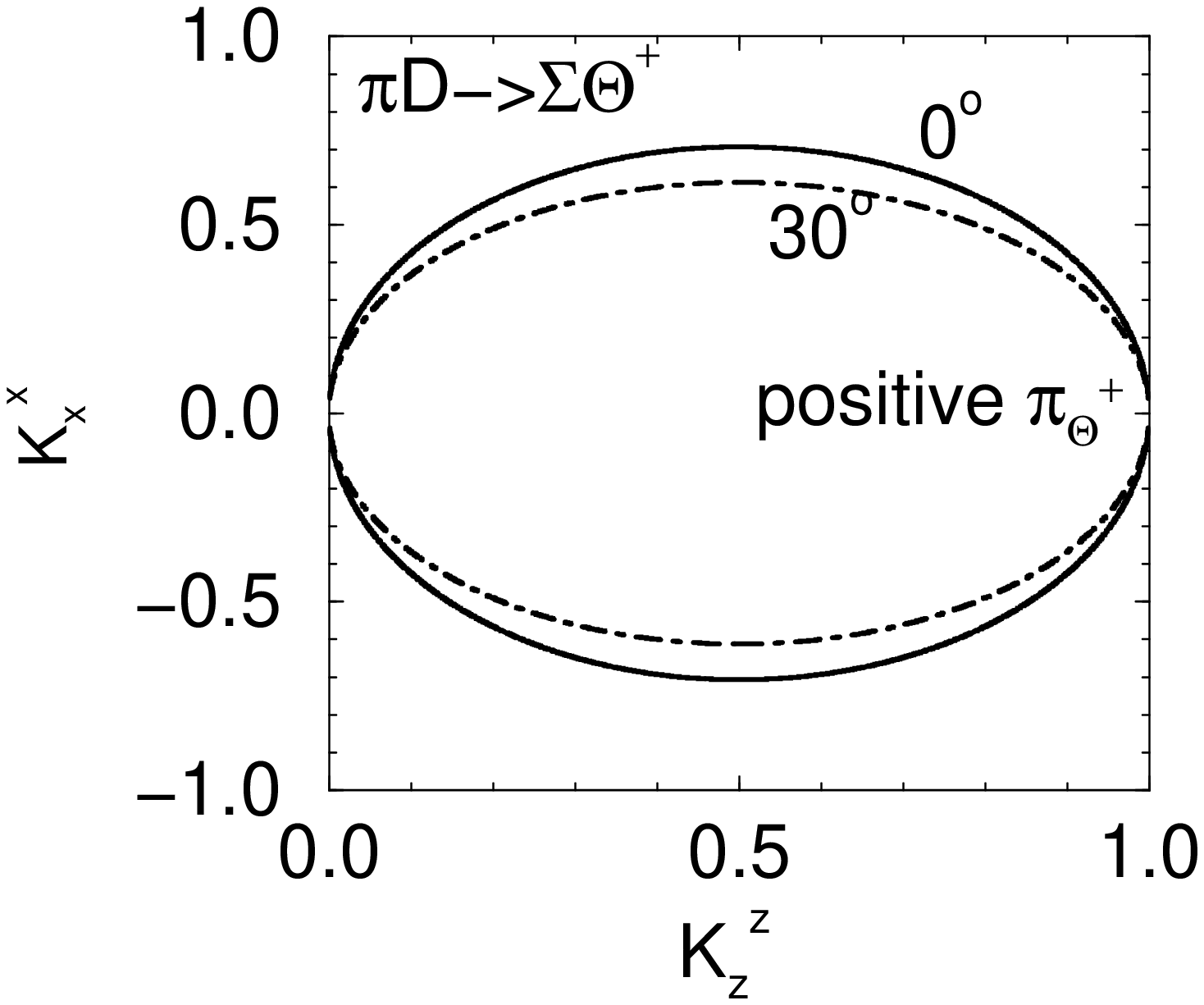}\qquad
  \includegraphics[width=.45\columnwidth]{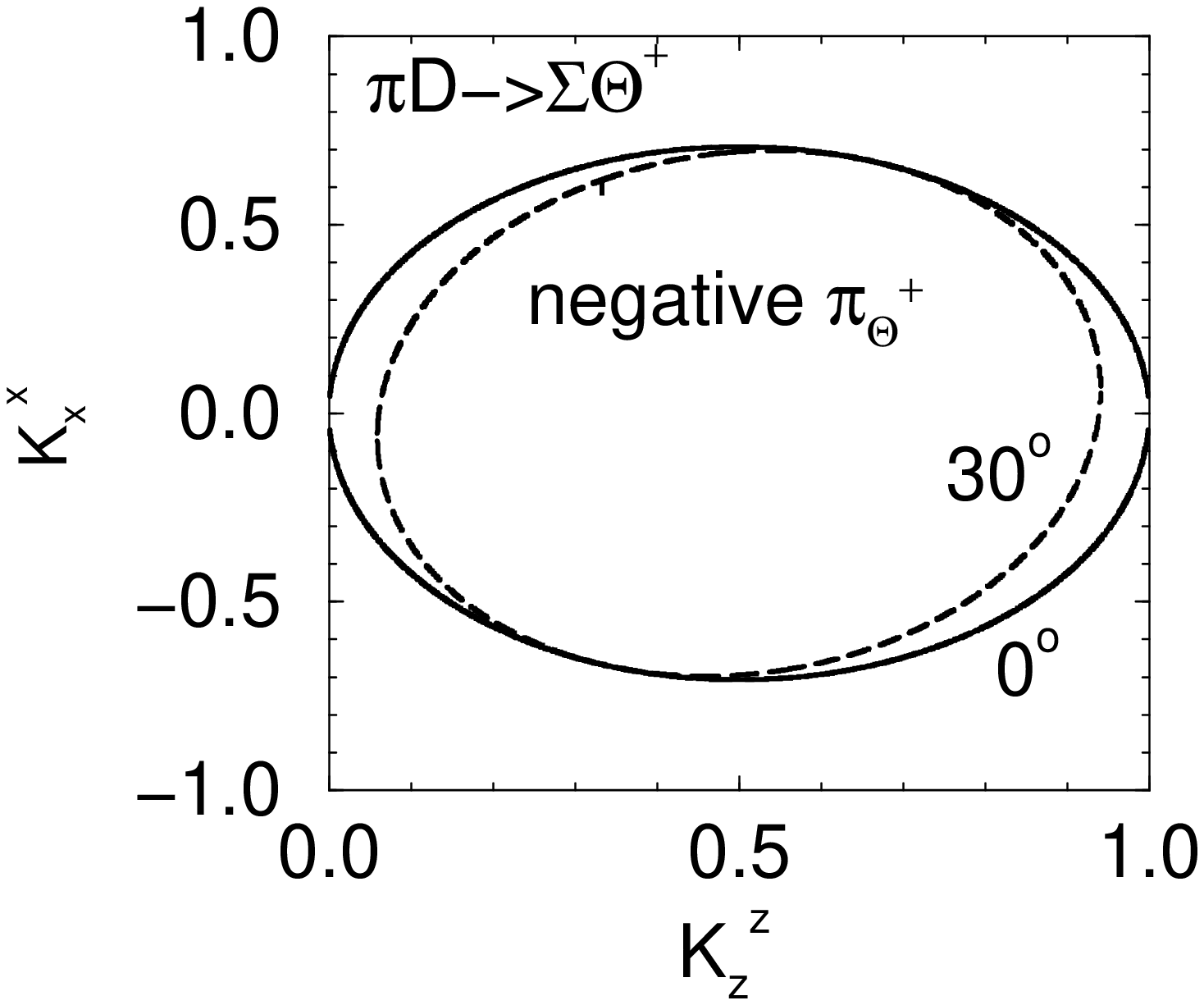}
 \caption{\label{fig:4}{\small%\tcaps%
 The spin-transfer coefficient $K_x^x$
 $\pi D \to\Sigma \Theta^+$ as a function of $K_z^z$ at two values
 of the relative phases $\delta^{\pm}=0^o,\, 30^0$ for
 positive (left panel) and negative (right panel) $\pi_\Theta$.}}}
\end{figure}

Fig.~\ref{fig:4} shows the dependence of the spin-transfer
coefficients $K_x^x$ on $K_z^z$ for two values of the relative
phases, $\delta^{\pm }=0^{o} $ and $30^{o} $ for positive (right
panel) and negative (left panel) $\pi_\Theta$. One can see a tiny
modification of $K_x^x$ when the phases $\delta^\pm$ become
finite.

\section{Deuteron spin anisotropy}

 Let us consider now the dependence of the probability of $\Sigma\Theta^+$  production
 as a function of the angle $\theta$ between deuteron spin polarization and
 the direction of the pion beam. For this aim we analyze the
 angular distribution $W(\cos\theta)$, normalized as
 \begin{eqnarray}
 \int\limits_{-1}^1 W(\cos\theta)\, d\cos\theta=1~.
 \label{E11}
 \end{eqnarray}
 This distribution may be found by using the general
 form of the production amplitude in Eq.~(\ref{E3}) and choosing
 the quantization axis along the deuteron spin. It has the universal
 form
 \begin{eqnarray}
 W^{\pm}(\cos\theta)= \frac{3}{2(3+ {\cal A}^{\pm})}
 \left( 1+{\cal A}^\pm \cos^2\theta\right)~,
 \label{E12}
 \end{eqnarray}
 where ${\cal A}^{\pm}$ is the deuteron spin anisotropy,
 and the superscript $\pm$
 indicates the $\Theta^+$ parity. The anisotropies may be
 expressed in explicit form through the partial amplitudes
 \begin{subequations}
 \label{E13}
 \begin{eqnarray}
 {\cal A}^+&=&\frac{3r^2-2}{3r^2 +2}~,\label{E13-1} \\
 {\cal A}^-&=&\frac{3(2\sqrt{2}\beta q - q^2 )}
 {4-2\sqrt{2}\beta q +5 q^2}~.\label{E13-2}
 \end{eqnarray}
 \end{subequations}
 Using  Eqs.~(\ref{E5}) and(\ref{E8}), we find the following relations
 \begin{eqnarray}
  {\cal A}^+={\cal A}^-&=&\frac{3K_z^z - 2}{2-K_z^z}~.
 \label{E14}
 \end{eqnarray}

\begin{figure}[t!]
{%\centering
  \includegraphics[width=.5\columnwidth]{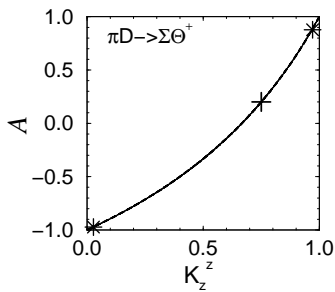}
%  \qquad
%  \includegraphics[width=.45\textwidth]{A_anisotopy1.eps}
 \caption{\label{fig:5}{\small%\tcaps%
 The deuteron spin-transfer anisotropy  $\cal A$
 as a function of $K_z^z$. The solutions for $r=q=1$
 for positive and negative $\pi_\Theta$
 are shown by cross and stars, respectively.}}}
 \end{figure}

 One can see that the asymmetry ${\cal A}$ as a function of $K_z^z$
 does not depend on $\pi_\Theta$.
 It is also important,  that the shape of the spin anisotropy as a
 function of $K_z^z$ does not depend on the phases $\delta^\pm$ and
 therefore is fully model independent. Again, in spite of the
 universality of the shape ${\cal A}(K_z^z)$ we find a strong
 difference in dependence of the anisotropy on the partial
 amplitudes for different parities. As an example, in Fig.~\ref{fig:5}
 we show solutions for positive and negative $\pi_\Theta$ at
 $r=q=1$.  One can see a strong difference in ${\cal A}$ for these
 different cases. For positive $\pi_\Theta$, ${\cal A}\simeq 0.2$, whereas
 for negative $\pi_\Theta$ we have ${\cal A}\simeq \pm1$, depending on
 phase of $\beta$.  This strong difference may be studied experimentally.
 % {\color{blue}{\it The
 % dependence on $\delta^+$ for the case of the positive $\pi_\Theta$
 % appears in dependence of the range of variation of $K_z^z$ (cf.
 % Eq.~(\ref{E6})).}}

\section{summary}

In summary, we have analyzed two types of spin observables in the
reaction $\pi \vec D\to\vec\Sigma\Theta^+$ near the threshold. One
type of observables concerns the spin-transfer coefficients
$K_x^x$ and $K_z^z$. Another one is the spin anisotropy. We found
a model independent correlation between the anisotropy and
spin-transfer coefficient $K_z^z$.  Each of these observables has
their own dependence on the partial amplitude and may be
calculated in a corresponding dynamical model which may be used
for the determination of the $\Theta^+$ parity.

 For the practical
implementation of the suggested measurements one needs a pion beam
at fixed energy impinging a polarized deuteron target and a large
phase space detector for the identifying $\Theta^+\to NK$ and
$\Sigma\to N\pi$ decay channels. The $\vec\Sigma^\pm$ polarization
is measurable in the standard way, by analyzing  the angular
distribution in the decay channel. The available secondary pion
beam delivered at the SIS/GSI Darmstadt and the spectrometers FOPI
and HADES make such a measurement feasible. As a first step one
can study the spin anisotropy which needs only a polarized
deuteron target but does not require a measurement of the spin of
outgoing particles.

\acknowledgments

We thank  H.W.~Barz, F.~Dohrmann, A.~Hosaka, L.P.~Kaptari,
R.~Kotte, K.~M\"oller, L.~Naumann, Yu.~Uzikov and S.~Zschocke  for
fruitful discussions. One of authors (A.I.T.) thanks E.~Grosse for
offering the hospitality at FZR.

\end{document}